\documentclass[letterpaper]{article} % DO NOT CHANGE THIS
\usepackage{aaai23} % \usepackage[submission]{aaai23}  % DO NOT CHANGE THIS
\usepackage{times}  % DO NOT CHANGE THIS
\usepackage{helvet}  % DO NOT CHANGE THIS
\usepackage{courier}  % DO NOT CHANGE THIS
\usepackage[hyphens]{url}  % DO NOT CHANGE THIS
\usepackage{graphicx} % DO NOT CHANGE THIS
\usepackage{natbib}  % DO NOT CHANGE THIS AND DO NOT ADD ANY OPTIONS TO IT
\usepackage{caption} % DO NOT CHANGE THIS AND DO NOT ADD ANY OPTIONS TO IT
\usepackage{algorithm}
\usepackage{algorithmic}

\usepackage{newfloat}
\usepackage{listings}
\DeclareCaptionStyle{ruled}{labelfont=normalfont,labelsep=colon,strut=off} % DO NOT CHANGE THIS
\floatstyle{ruled}
\newfloat{listing}{tb}{lst}{}
\floatname{listing}{Listing}
\nocopyright% -- Your paper will not be published if you use this command
\title{A Framework for Modeling, Analyzing, and Decision-Making in Disease Spread Dynamics and Medicine/Vaccine Distribution}
\author {
    Zenin Easa Panthakkalakath,\textsuperscript{\rm 1, \rm 2}
    Neeraj, \textsuperscript{\rm 2}
    Jimson Mathew \textsuperscript{\rm 2}
}
\affiliations {
    \textsuperscript{\rm 1} Università della Svizzera italiana\\
    \textsuperscript{\rm 2} Indian Institute of Technology Patna\\
    zenin.easa.panthakkalakath@usi.ch, neeraj.pcs17@iitp.ac.in, jimson@iitp.ac.in
}

\usepackage{bibentry}

\begin{document}

\maketitle

\begin{abstract}
    The challenges posed by epidemics and pandemics are immense, especially if the causes are novel. This article introduces a versatile open-source simulation framework designed to model intricate dynamics of infectious diseases across diverse population centres. Taking inspiration from historical precedents such as the Spanish flu and COVID-19, and geographical economic theories such as Central place theory, the simulation integrates agent-based modelling to depict the movement and interactions of individuals within different settlement hierarchies. Additionally, the framework provides a tool for decision-makers to assess and strategize optimal distribution plans for limited resources like vaccines or cures as well as to impose mobility restrictions.
\end{abstract}

%%%%%%%%%%%%%%%%%%%%%%%%%%%%%%%%%%%%%%%%%%%%%%%%%%%%%%%%%%%%%%%%%%%%%%%%%%%%%%%%
\section{Introduction}

A century ago, the Spanish flu \cite{trilla20081918} infected over 500 million. One-third of the world population at that time - in four successive waves in about two years. Now, we have yet another viral infection that has taken over the world by storm. The outbreak of COVID-19 was first confirmed in Wuhan, China, in December 2019 \cite{sohrabi2020world}. The World Health Organization declared a global public health emergency on 30 January 2020. The pandemic has affected the health of the people and the economy at large, which makes this a huge crisis.

An article in Washington post, titled 'Why outbreaks like coronavirus spread exponentially, and how to "flatten the curve"', written by Harry Stevens \cite{stevens2020outbreaks}. The article talks about the mathematical modeling of how diseases spread and the importance of social distancing. The article shows the exponential nature of disease spread using simple animation. It follows an Agent-Based Modelling approach for the same. There have been a few attempts to recreate similar modeling by others in the scientific community. A renowned Mathematician and MATLAB creator, Cleve Moler had created his own version of the COVID-19 simulator in MATLAB and published in his blog \cite{C2020Moler}. Similar attempts to reproduce the same using Simulink and Stateflow can be seen in a blog post by Guy Rouleau \cite{G2020Rouleau}.

Studies indicate that there has been a massive decrease in the number of people in the central districts of cities due to the pandemic \cite{S2020Lehmann}. It can be interpreted that the reason for such a decline is because there have been a huge number of people who migrated into such centers for economic benefit, now returning to their native region. It is safe to assume that an overwhelming majority of these people come from lower-order settlements like smaller towns and villages around these cities.

We do not know for certain if a vaccine or a cure would be available soon. Even so, we must keep being optimistic and trust the scientific community that we will have our hands on either or both of these soon. In this article, we are assuming that a vaccine would provide immunity to an uninfected person. A cure would enable a sick person to be free of a disease without making the person immune to the disease.

Once such a vaccine or a cure is developed, an administrative decision needs to be taken regarding how the vaccine must be distributed. It is not far-fetched to think that such a vaccine or cure would only be available in limited numbers as the manufacturing facilities would be limited. With the limited amount of resources at hand, the administration bodies must attempt to make optimal decisions to minimize the disease's fatalities. Various parameters would affect such a decision-making process. Some of these are the population density, the number and percentage of infected people, the number and percentage of people who had the disease and recovered, the mobility of the people, quantity of the vaccine or cure available at a given time and the frequency at which the vaccine or cure can be procured.

This article presents a simulation framework that models an environment containing different boundaries and agents moving in the environment. The boundaries depict different settlement hierarchy limits, such as city, town, or village boundaries. The boundaries may or may not be permeable. The agents moving in the environment depict the region's population and have properties to depict whether they are infected, recovered/immune, or uninfected. The framework is parameterized so that the positioning of the boundaries and their impermeability can be changed. Also, the size, location, and the initial number of infected people in a population can be set before the start of the simulation. It is suggested that the choice of modeling method is made based on decision-makers' requirements, type of problem and system complexity, and its characteristics, which could also be derived from economic models \cite{tako2018comparing, tako2012application}, Borshchev and Filippov \cite{borshchev2004system},  Brennan et al. \cite{brennan2006taxonomy}, Chahal and Eldabi \cite{chahal2010multi}, McHaney et al. \cite{mchaney2018using}.

%%%%%%%%%%%%%%%%%%%%%%%%%%%%%%%%%%%%%%%%%%%%%%%%%%%%%%%%%%%%%%%%%%%%%%%%%%%%%%%%
\section{Preliminaries}

%%%%%%%%%%%%%%%%%%%%%%%%%%%%%%%%
\subsection{Central Place Theory}
Introduced in 1933, Central Place Theory is a theory \cite{berry1958recent} that attempts to explain the spatial distribution of human settlements. The theory was originally analyzed by a German geographer named Walter Christaller. This theory was the foundation of cities' study as cities' systems, rather than simple hierarchies or single entities. The theory is based on various assumptions. All regions have an isotropic surface with evenly distributed population and resources, having similar cost and quality for goods and services, and consumers having similar purchasing power.

The theory proposes a hexagonal grid (or honeycomb) structure of distribution of cities, towns, and villages. Effectively, there shall be higher-order settlements surrounded by lower-order settlements in the grid. Each higher-order settlements shall be equidistant from other higher-order settlements, and so is the case with lower-order settlements.

%%%%%%%%%%%%%%%%%%%%%%%%%%%%%%%%
\subsection{Agent-Based Modeling}
Agent-Based Modeling is a style of modeling for simulating autonomous agents' actions and interactions in an environment to predict the emergence of complex group behaviors. It is a kind of microscale model that enables simulation of multiple agents performing simultaneous operations and interacting between themselves and the environment in which they reside in. The simulation may also involve different decision-making heuristics and learning rules or adaptive processes. In an analysis, Epstein \cite{epstein2009modelling} suggests that ABM is appropriate for modeling pandemics.

The origin of the idea of agent-based modeling can be traced back to the Von Neumann Machine, which was designed in the 1940s. It was a theoretical self-replicating machine in a cellular automata environment. This method has recently been used to model and simulates various scenarios, including, but not limited to, automated driving, cognitive, social simulation, satellite constellation modeling, multi-robot systems, etc.

%%%%%%%%%%%%%%%%%%%%%%%%%%%%%%%%%%%%%%%%%%%%%%%%%%%%%%%%%%%%%%%%%%%%%%%%%%%%%%%%
\section{Framework Development}

Our goal is to create a framework that people can use to extend it to model disease spreading scenarios and distribution of vaccines and/or medicine. During the development process, several assumptions were made, and only some features were prioritized.

%%%%%%%%%%%%%%%%%%%%%%%%%%%%%%%%
\subsection{Assumptions}

The following assumptions were made while developing the framework. This was done in order to simplify the development process.

\begin{enumerate}
    \item All persons in the model are identical. Age, immunity level, or any other factors are not considered. The fatality of a person is randomly determined during the simulation.
    \item A person may move freely in accordance with a randomly assigned velocity and direction until it encounters the boundary. The velocity is assigned during the start of the simulation and the magnitude remains unchanged.
    \item As a person encounters the boundary, the person may either cross the boundary or get reflected from the boundary. Whether a person bounces off or crosses the boundary is determined in accordance with the probability value assigned to each boundary in the setup.
    \item A person may acquire the disease when close to an infected person while within a particular distance from the infected person, and in accordance with the transmission probability.
    \item A person who gets cured of the disease without having the medicine shall become immune.
    \item An uninfected person who receives the vaccine becomes immune.
    \item Only an uninfected person may receive the vaccine.
    \item An infected person who receives the medicine becomes equivalent to be uninfected. The person shall not become immune.
    \item Only an infected person may receive the medicine.
    \item A person who is infected may die in accordance with a probability value.
    \item A person may get cured after being infected for a fixed period of time (or iterations).
\end{enumerate}

%%%%%%%%%%%%%%%%%%%%%%%%%%%%%%%%
\subsection{Features}

The key features supported by the framework are:

\begin{enumerate}
    \item Ability to draw straight line boundaries at different orientations that may be permeable, impermeable, or partial permeable. The user can set a value for the probability of whether a person encountering the boundary would be reflected or transmitted.
    \item Ability to define a region inside which people (as agents) can be generated. The center coordinate and the radius of the region need to be specified along with the number of people that need to be generated within that region and the number of people in that who are infected.
    \item Ability to specify parameters related to disease spread radius, probability of disease transmission for a person within spread radius of an infected person, probability of getting killed at any time-step while being sick and time period to be free from being sick.
    \item Ability to specify parameters related to lockdown implementation. There are parameters to specify at what threshold the ratio of the number of sick people to the total population lockdown must be implemented and revoked. While under lockdown, the people's mobility under lockdown would be multiplied by a lockdown mobility multiplier parameter, which assumes a value between 0 and 1.
    \item Ability to distribute vaccine and/or medicines with parameters to specify the quantity, the start time-step, the frequency at which these get replenished and the mode of distribution. The framework supports four modes of vaccine and medicine distribution based on the statistical state of the agents in the region. These are:
        \begin{enumerate}
            \item "equitable": Distribution proportional to the number of uninfected people in the region at the given time.
            \item  "maximumInfection": Distribution prioritizing regions with maximum number of infected cases first.
            \item  "maximumUninfected": Distribution prioritizing regions with maximum number of uninfected people first.
            \item  "infectedAndUninfected": Distribution prioritizing regions with maximum number of sum of uninfected and infected people first.
        \end{enumerate}
\end{enumerate}

%%%%%%%%%%%%%%%%%%%%%%%%%%%%%%%%
\subsection{Implementation}

The framework is implemented in JavaScript with HTML and CSS to support the visualization. It can load all the simulation parameters from a JSON file. These parameters include everything related to boundaries and their orientation, region generation and population information, disease transmission, lockdown implementation, and vaccine and medicine distribution.

The framework loads the boundaries in the orientation specified by the endpoints by reflecting on the boundary parameters specified in the aforementioned JSON file. The line's color is in accordance with the value of the boundary's impermeability and the value for the angle of orientation. During the simulation, the agents read the boundary's color when encountered to compute the impermeability value. If it is decided that the agent is to be reflected from the boundary, the orientation angle comes in handy to calculate the direction of motion post reflection.

People are the agents in this simulation. Uninfected people are represented with a black dot. The color changes to red when they get infect and to green when they recover or get immune. People who die are deleted from the simulation from that point forward.

%%%%%%%%%%%%%%%%%%%%%%%%%%%%%%%%%%%%%%%%%%%%%%%%%%%%%%%%%%%%%%%%%%%%%%%%%%%%%%%%
\section{Simulation Methodology}

Using the framework, we can set up different scenarios involving different kinds of borders according to area layouts and people's simple behavior in different regions. A user of this framework trying to model a city or town of his/her interest can decide on the boundaries and other parameters.

Once the environment model is ready, the user can try simulating without adding any vaccine or cure into the simulation and observe if the data output is similar to what the actual city has been providing in the past many months. If not, the user can keep tweaking the parameters until the values come close to the actual data.

Now that the model is behaving similar to the real-world scenario, we can add in vaccines and/or cure to the equation and run different simulations and observe the results. Try to find the decisions involving the least amount of fatality, which is the decision that the administration must make regarding vaccine distribution.

Below mentioned are the steps that one must follow to use the framework to model for their region.

\begin{enumerate}
    \item Create a new JSON file named 'myRegion.json'. Edit the variable named 'environmentJSONFile' in 'script.js' file to point to this file.
    \item Start by drawing the boundaries. Create a variable named "boundaries", which is an array of arrays. Each child array represents a boundary, with each of its elements being X-coordinate of the first point, Y-coordinate of the first point, X-coordinate of the second point, Y-coordinate of the second point, and impermeability. The value of impermeability ranges between 0 and 1, 0 means completely permeable, and 1 means completely impermeable.
    \item Define the region inside which persons get generated. Add in a variable named "regions," which is an array of objects. Each object corresponds to a circular region inside which the specified number of people gets generated. The properties of the object are "id", "population", "infected", "center", "radius" and "mobilityFactor", which corresponds to a string identity of the region, total population in the region, number of infected people in the region, geographic center coordinate of the region (an array having two values corresponding to X-coordinate and Y-coordinate), the radius of the region and mobility factor (which corresponds to the velocity of the people in the simulation).
    \item Finally, define other environmental parameters "boundaryThickness", "spreadRadius", "curePeriod" and "killProbability". Environmental parameters like "lockdownStartThreshold", "lockdownEndThreshold" and "lockdownMobilityMultiplier". These may be chosen in accordance with how the Government usually reacts to the situation. Environmental parameters related to vaccine and medicine distribution may be set to a value such that it wouldn't make any impact, say we set the quantity to zero.
    \item Run the simulation, download the result, and observe whether the simulation results are coherent with the disease spread information for the city at hand. If not, check and modify the environmental parameters until the data is coherent enough.
    \item Now, modify the parameters related to vaccine and medicine distribution in accordance with the potential future availability of the same.
\end{enumerate}

To showcase the framework's capability, we have come up with a region layout involving hexagonal grids. The hexagons in the grid would represent villages, towns, or cities. The Central Place Theory inspires this layout and the modeling of many cities may be similar to this.

\begin{figure}[htb!]
    \centering
    \includegraphics[width=0.45\textwidth]{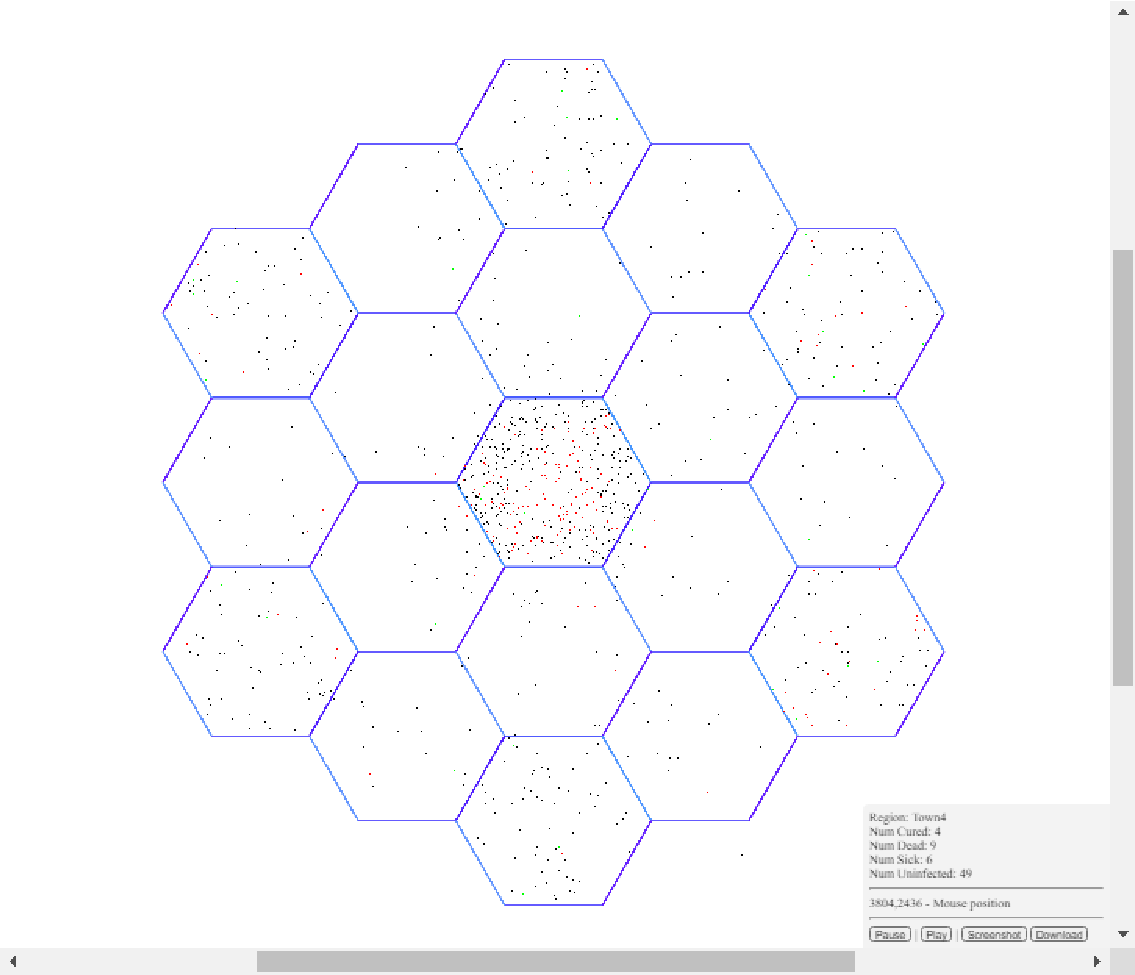}
    \caption{A screenshot showcasing the hexagonal grid (honeycomb) layout depicting different settlement hierarchies and dots depicting the population in the region.}
    \label{fig:simulationStartScreenshot}
\end{figure}

\begin{figure}[htb!]
    \centering
    \includegraphics[width=0.45\textwidth]{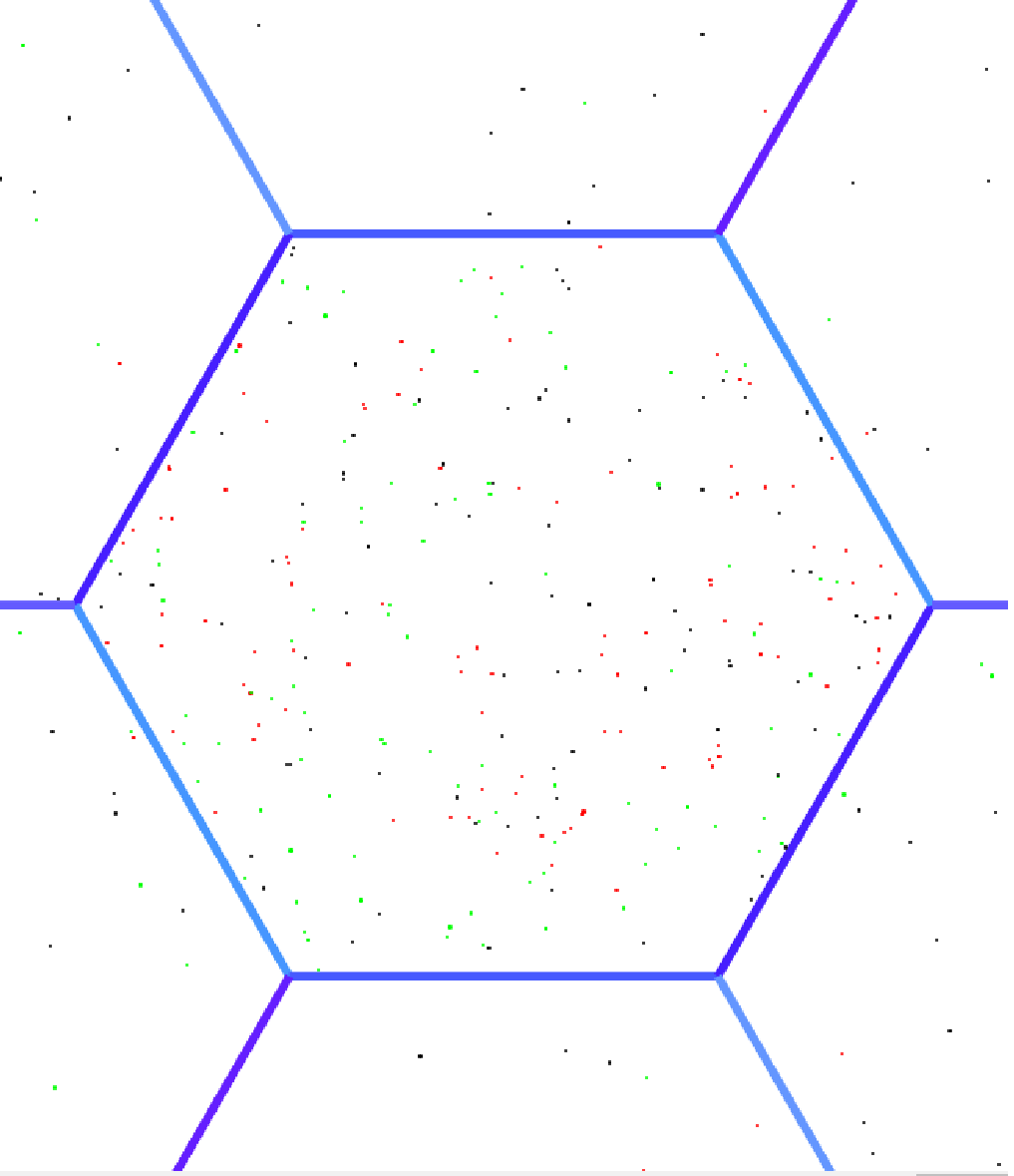}
    \caption{A screenshot focusing the central city location. Here we can clearly see the dots with different colors. Black depicts people who are uninfected, red depicts people who are currently infected, and green depicts people who have recovered from the disease and are immune to it.}
    \label{fig:centeralCityScreenshot}
\end{figure}

The hexagonal grids were programmatically generated and the values for boundary impermeability and the values for initial population and infected people in each region were entered manually. As explained in the aforementioned points, the parameters were tweaked until a reasonable simulation result was observed with the values for vaccine and medicine quantities set to zero.

In all the scenarios that we are trying to simulate, we are keeping some parameters constant. The values for these parameters can be seen in Table \ref{tab:hyperparams}.

\begin{table}[htb!]
\centering
\begin{tabular}{|c|c|}
\hline
\textbf{Parameter name} & \textbf{Value}  \\ \hline
spreadRadius & 5 \\ \hline
curePeriod & 250 \\ \hline
killProbability  & 0.005  \\ \hline
transmissionProbability & 0.7 \\ \hline
\end{tabular}
\caption{Parameters used in the simulation that were kept constant across various scenarios.}
\label{tab:hyperparams}
\end{table}

\begin{figure}[htb!]
    \centering
    \includegraphics[width=0.45\textwidth]{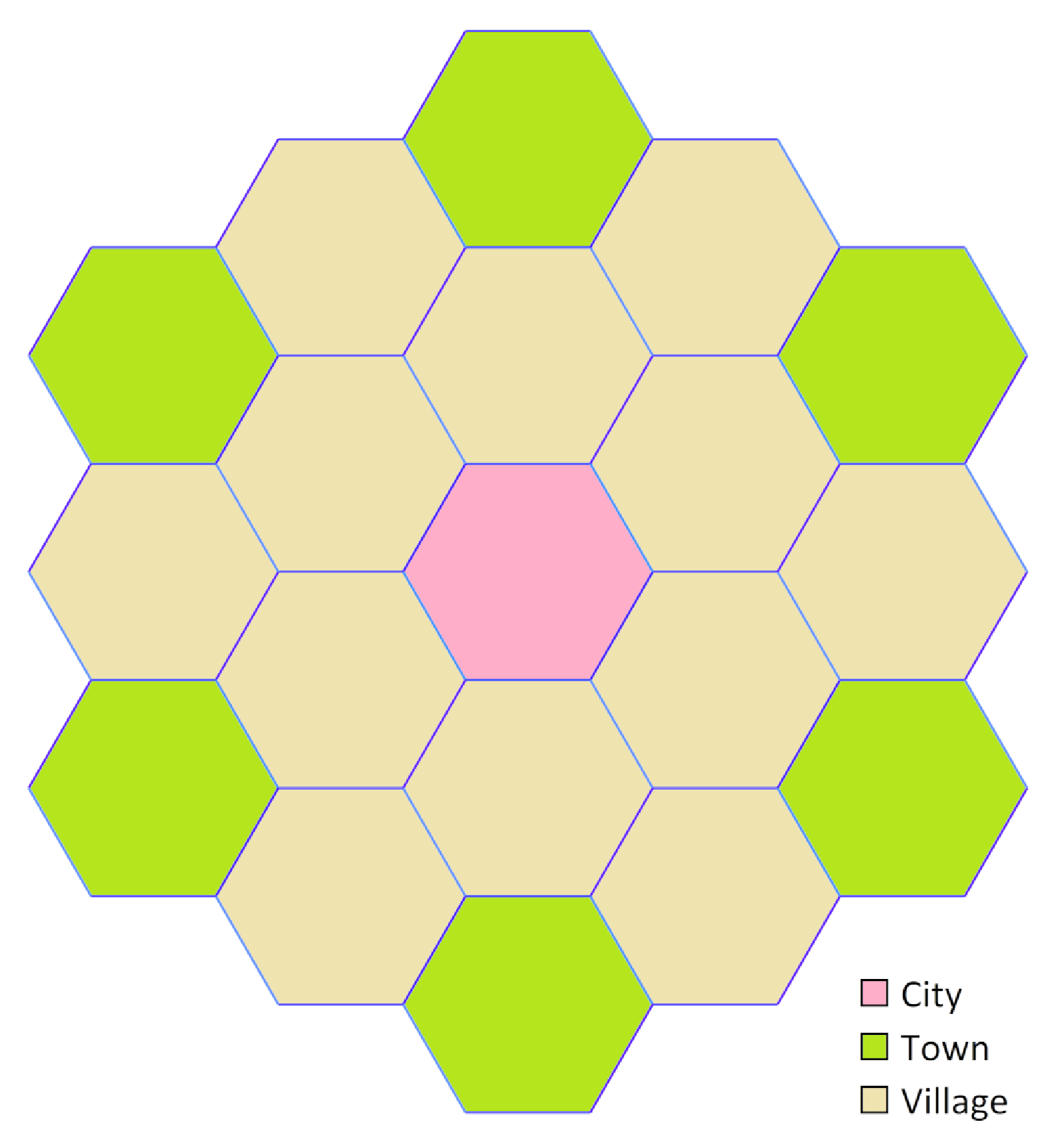}
    \caption{An image depicting the placement of city, towns and villages in a hexagonal grid.}
    \label{fig:citytownvillageplacement}
\end{figure}

In the hexagonal grid, the city, towns, and villages were placed, as shown in Figure \ref{fig:citytownvillageplacement}. The total population, the number of infected people in each region type, and the people's mobility factor in these regions are mentioned in Table \ref{tab:hierpopandmob}.

\begin{table}[htb!]
\centering
\begin{tabular}{|c|c|c|c|}
\hline
\textbf{Hierarchy} & \textbf{Population} & \textbf{Infected} & \textbf{Mobility}  \\ \hline
City & 750 & 5 & 1 \\ \hline
Town & 100 & 5 & 3 \\ \hline
Village & 15 & 1 & 5 \\ \hline
\end{tabular}
\caption{Initial conditions for different settlement hierarchies assumed in the simulation. The values may be considered scaled.}
\label{tab:hierpopandmob}
\end{table}

The value of the impermeability of the borders were decided based on who shares the border. Even during a pandemic, the border between a city and a village would be more permeable compared to the the border between a village and a town, which would be more permeable that the border between two villages. This is because people from villages would be dependent on nearby towns and cities for economic activities. Even during a pandemic situation, such an activity would be required to ensure survival of the population. The values of the same that we adopted for this simulation is shown in Table \ref{tab:borderimpermeability}.

\begin{table}[htb!]
\centering
\begin{tabular}{|c|c|c|c|}
\hline
\textbf{Between hierarchies} & \textbf{Impermeability}  \\ \hline
City-Village & 0.7 \\ \hline
Town-Village & 0.8 \\ \hline
Village-Village & 1 \\ \hline
\end{tabular}
\caption{The border impermeability values between different settlement hierarchy.}
\label{tab:borderimpermeability}
\end{table}

Arranging cities, villages, and towns in such an arrangement enables capturing the phenomenon of population migration from higher orders of settlements to lower orders of settlements during a pandemic time. It is also important to differentiate the central city's statistical data to that of the entire region. A correlation between these two sets of statistics is important, as a survey that doesn't consider such a migration would have missing links.

%%%%%%%%%%%%%%%%%%%%%%%%%%%%%%%%%%%%%%%%%%%%%%%%%%%%%%%%%%%%%%%%%%%%%%%%%%%%%%%%
\section{Results}

After the development, the framework was put to test to observe the simulation results under various circumstances. These are described in the following subsections. Note that the simulation was conducted for a hypothetical hexagonally divided regions having population scaled to smaller numbers.

%%%%%%%%%%%%%%%%%%%%%%%%%%%%%%%%
\subsection{Simulation Results without Vaccine, Cure or Lockdown}

\begin{figure}[htb!]
    \centering
    \includegraphics[width=0.45\textwidth]{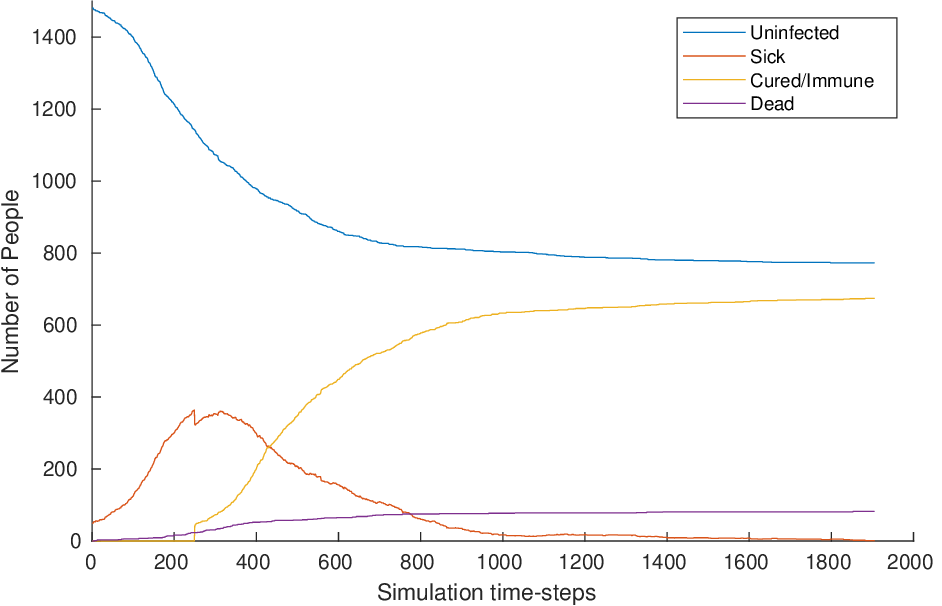}
    \caption{The aggregate simulation results for all the regions when no vaccine, medicine or lockdown were implemented.}
    \label{fig:allregionsnone}
\end{figure}

\begin{figure}[htb!]
    \centering
    \includegraphics[width=0.45\textwidth]{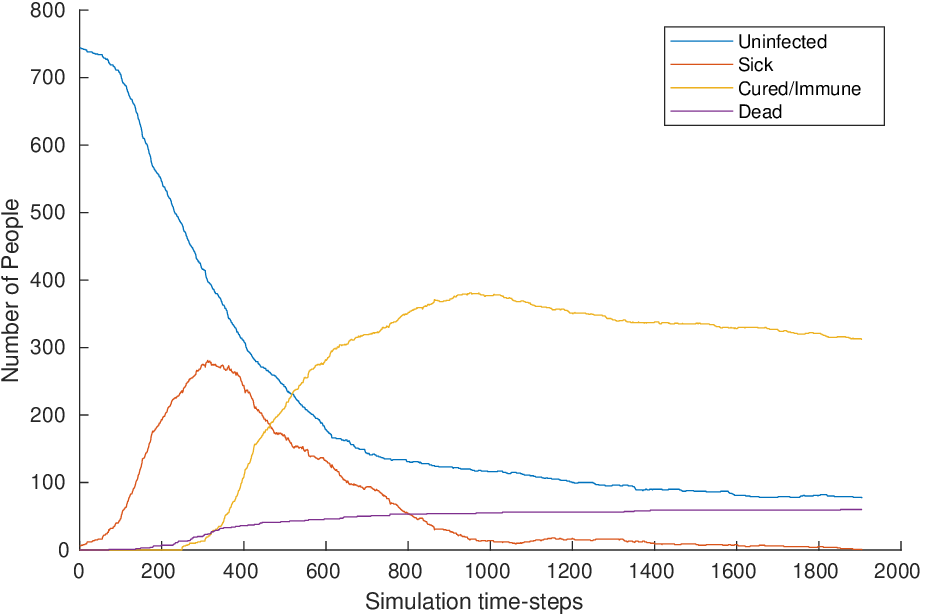}
    \caption{The simulation results for the central city when no vaccine, medicine or lockdown were implemented.}
    \label{fig:centralcitynone}
\end{figure}

If no restrictions are imposed by the administrative authorities, the scenario would end up something similar to what is depicted in Figures \ref{fig:allregionsnone} and \ref{fig:centralcitynone}.


At the end of the simulation, the statistics of the number of people who where dead and recovered are mentioned in Table \ref{tab:noneresult}.

\begin{table}[htb!]
\centering
\begin{tabular}{|c|c|}
\hline
\textbf{Parameter name} & \textbf{Value}  \\ \hline
Initial Population & 1530 \\ \hline
Simulation Period & 1905 \\ \hline
Total Immune & 675 \\ \hline
Total Dead & 82 \\ \hline
\end{tabular}
\caption{Tabulated final results for all the regions when no vaccine, medicine or lockdown were implemented.}
\label{tab:noneresult}
\end{table}

%%%%%%%%%%%%%%%%%%%%%%%%%%%%%%%%
\subsection{Simulation Results with lockdown}

Simulation was conducted wherein lockdown was implemented whenever more than 10\% were infected and revoked whenever less than 2\% were infected in any particular region. The parameter values for the same is described in Table \ref{tab:lockdownparams}.

\begin{table}[htb!]
\centering
\begin{tabular}{|c|c|}
\hline
\textbf{Parameter name} & \textbf{Value}  \\ \hline
lockdownStartThreshold & 0.1 \\ \hline
lockdownEndThreshold & 0.02 \\ \hline
lockdownMobilityMultiplier & 0.1 \\ \hline
\end{tabular}
\caption{Simulation parameters used to implement lockdown.}
\label{tab:lockdownparams}
\end{table}

\begin{figure}[htb!]
    \centering
    \includegraphics[width=0.45\textwidth]{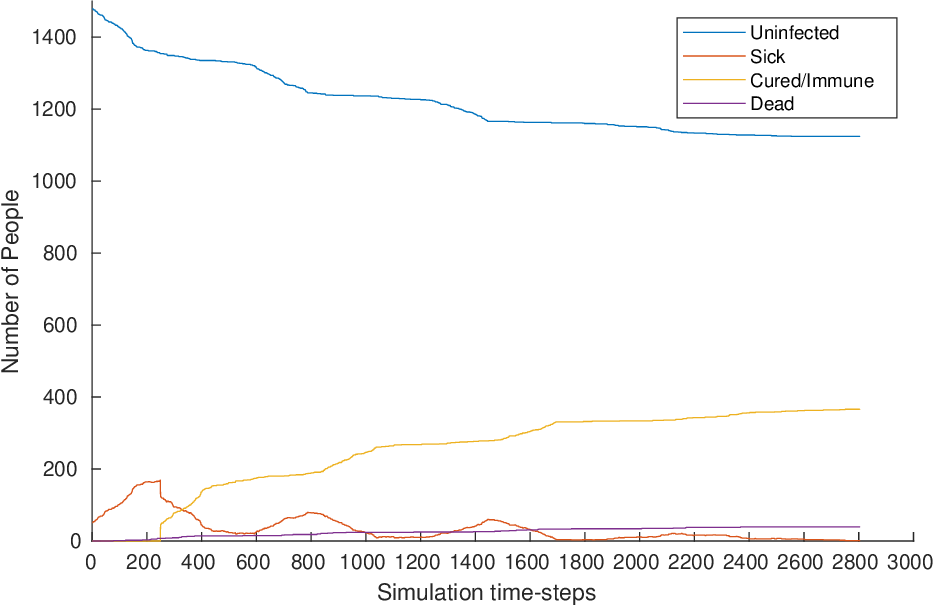}
    \caption{The aggregate simulation results for all the regions when lockdown was implemented.}
    \label{fig:allregionsWithLockdown}
\end{figure}
\begin{figure}[htb!]
    \centering
    \includegraphics[width=0.45\textwidth]{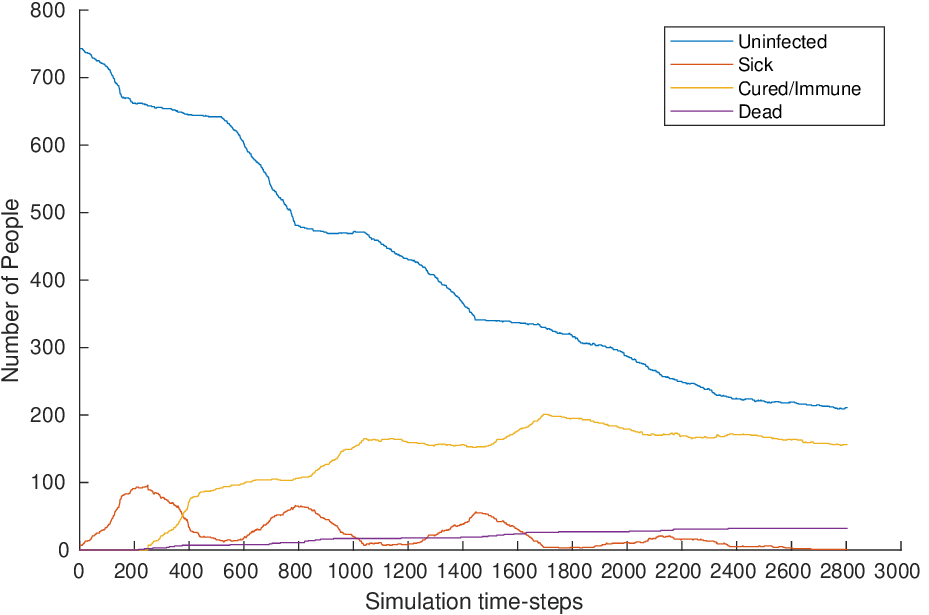}
    \caption{The simulation results for the central city when lockdown was implemented.}
    \label{fig:centralcityWithLockdown}
\end{figure}

At the end of the simulation, the statistics of the number of people who where dead and recovered are mentioned in Table \ref{tab:lockdownresult}.

\begin{table}[htb!]
\centering
\begin{tabular}{|c|c|}
\hline
\textbf{Name} & \textbf{Value}  \\ \hline
Initial Population & 1530 \\ \hline
Simulation Period & 2804 \\ \hline
Total Immune & 367 \\ \hline
Total Dead & 39 \\ \hline
\end{tabular}
\caption{Tabulated final results for all the regions with lockdown.}
\label{tab:lockdownresult}
\end{table}

We can observe that there would be four waves when a lockdown is implemented in such a manner.

%%%%%%%%%%%%%%%%%%%%%%%%%%%%%%%%
\subsection{Simulation Results with Vaccine distribution}

Simulations were conducted with all four supported methods of vaccine distribution. These are mentioned in the subsequent paragraphs, figures and tables. In all the cases, vaccine distribution starts at 300 time-step with a quantity of 200 units and the quantity gets replenished after every 100 subsequent time-step.

Firstly, let us observe the results when vaccine is distributed equitably i.e., proportional to the number of uninfected people in the region at the given time. The parameter values for the same is described in Table \ref{tab:vaccineEqutablyparams}.

\begin{table}[htb!]
\centering
\begin{tabular}{|c|c|}
\hline
\textbf{Parameter name} & \textbf{Value}  \\ \hline
vaccineDistributionStartTime & 300 \\ \hline
vaccineDistributionFrequency & 100 \\ \hline
vaccineDistributionQuantity & 200 \\ \hline
vaccineDistributionMechanism & "equitable" \\ \hline
\end{tabular}
\caption{Simulation parameters used when vaccine distribution mode was "equitable".}
\label{tab:vaccineEqutablyparams}
\end{table}

\begin{figure}[htb!]
    \centering
    \includegraphics[width=0.45\textwidth]{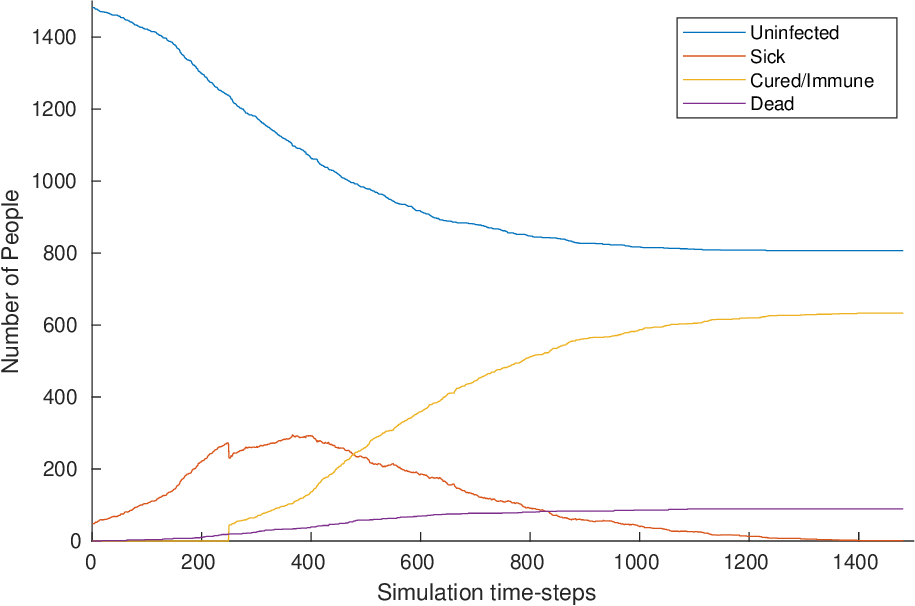}
    \caption{The aggregate simulation results for all the regions when vaccine distribution mode was "equitable".}
    \label{fig:allregionsWithVaccineEqutably}
\end{figure}
\begin{figure}[htb!]
    \centering
    \includegraphics[width=0.45\textwidth]{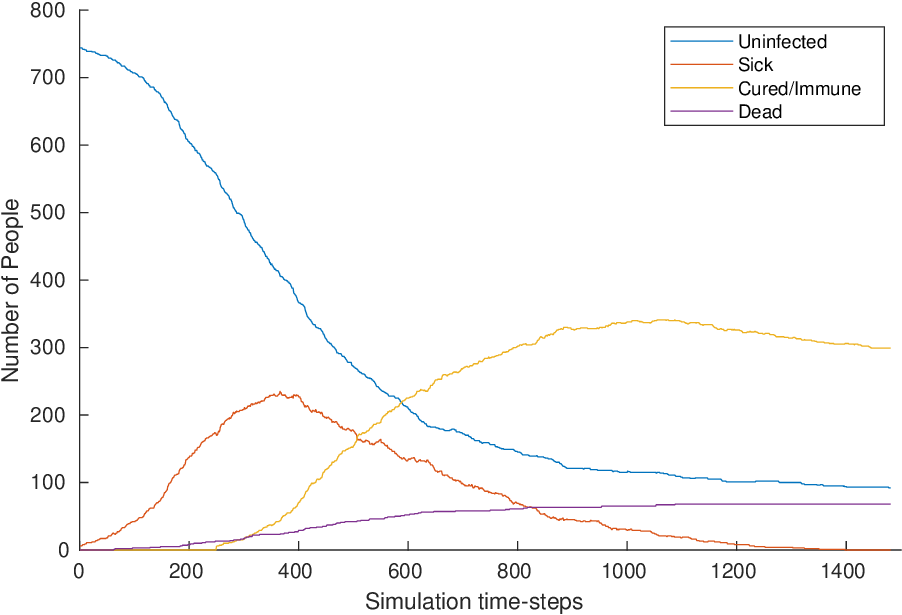}
    \caption{The simulation results for the central city when vaccine distribution mode was "equitable".}
    \label{fig:centralcityWithVaccineEqutably}
\end{figure}

At the end of the simulation, the statistics of the number of people who where dead and recovered are mentioned in Table \ref{tab:vaccineEqutablyResult}.

\begin{table}[htb!]
\centering
\begin{tabular}{|c|c|}
\hline
\textbf{Name} & \textbf{Value}  \\ \hline
Initial Population & 1530 \\ \hline
Simulation Period & 1481 \\ \hline
Total Immune & 634 \\ \hline
Total Dead & 89 \\ \hline
\end{tabular}
\caption{Tabulated final results for all the regions with vaccine distribution mode "equitable".}
\label{tab:vaccineEqutablyResult}
\end{table}

Secondly, let us observe the results when vaccine is distributed prioritizing regions with maximum number of infected cases first. The parameter values for the same is described in Table \ref{tab:vaccineMaximumInfectionParams}.

\begin{table}[htb!]
\centering
\begin{tabular}{|c|c|}
\hline
\textbf{Parameter name} & \textbf{Value}  \\ \hline
vaccineDistributionStartTime & 300 \\ \hline
vaccineDistributionFrequency & 100 \\ \hline
vaccineDistributionQuantity & 200 \\ \hline
vaccineDistributionMechanism & "maximumInfection" \\ \hline
\end{tabular}
\caption{Simulation parameters used when vaccine distribution mode was "maximumInfection".}
\label{tab:vaccineMaximumInfectionParams}
\end{table}

\begin{figure}[htb!]
    \centering
    \includegraphics[width=0.45\textwidth]{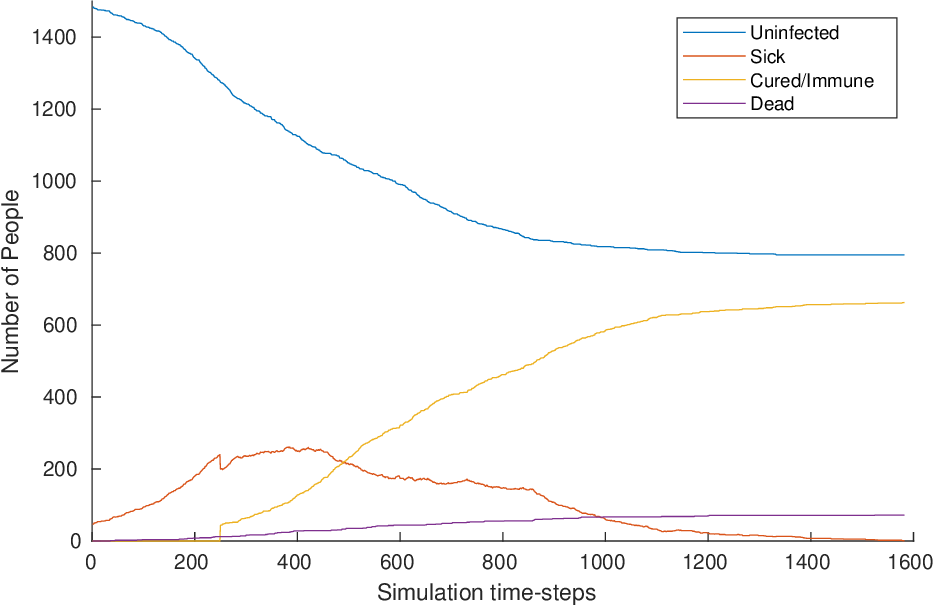}
    \caption{The simulation results for all the regions when vaccine distribution mode was "maximumInfection".}
    \label{fig:allregionsWithVaccineMaximumInfection}
\end{figure}
\begin{figure}[H]
    \centering
    \includegraphics[width=0.45\textwidth]{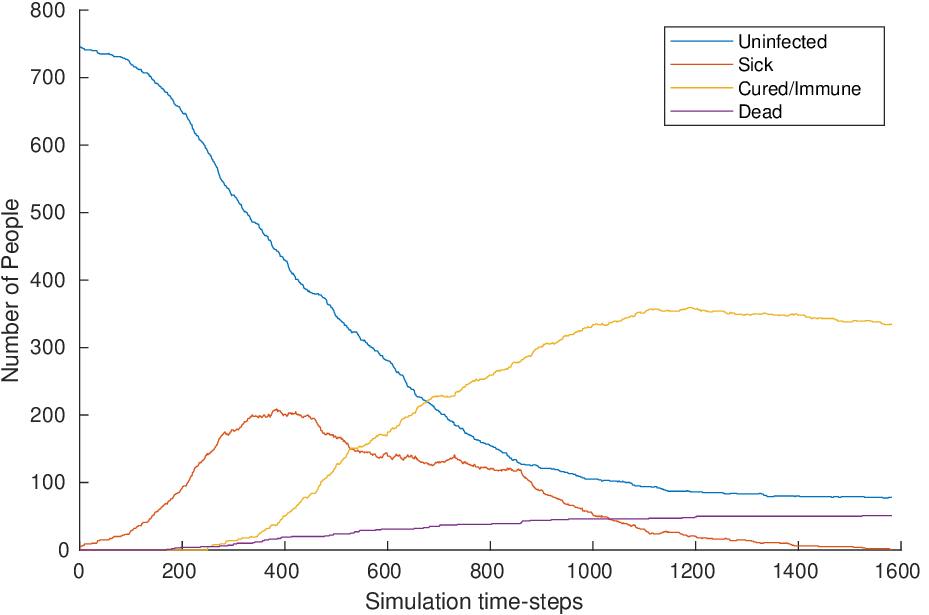}
    \caption{The simulation results for the central city when vaccine distribution mode was "maximumInfection".}
    \label{fig:centralcityWithVaccineMaximumInfection}
\end{figure}

At the end of the simulation, the statistics of the number of people who where dead and recovered are mentioned in Table \ref{tab:VaccineMaximumInfectionResults}.

\begin{table}[htb!]
\centering
\begin{tabular}{|c|c|}
\hline
\textbf{Name} & \textbf{Value}  \\ \hline
Initial Population & 1530 \\ \hline
Simulation Period & 1481 \\ \hline
Total Immune & 663 \\ \hline
Total Dead & 72 \\ \hline
\end{tabular}
\caption{Tabulated final results for all the regions with vaccine distribution mode "maximumInfection".}
\label{tab:VaccineMaximumInfectionResults}
\end{table}

Thirdly, let us observe the results when vaccine is distributed prioritizing regions with maximum number of uninfected people first. The parameter values for the same is described in Table \ref{tab:vaccineMaximumUninfectedParams}.

\begin{table}[htb!]
\centering
\begin{tabular}{|c|c|}
\hline
\textbf{Parameter name} & \textbf{Value}  \\ \hline
vaccineDistributionStartTime & 300 \\ \hline
vaccineDistributionFrequency & 100 \\ \hline
vaccineDistributionQuantity & 200 \\ \hline
vaccineDistributionMechanism & "maximumUninfected" \\ \hline
\end{tabular}
\caption{Simulation parameters used when vaccine distribution mode was "maximumUninfected".}
\label{tab:vaccineMaximumUninfectedParams}
\end{table}

\begin{figure}[htb!]
    \centering
    \includegraphics[width=0.45\textwidth]{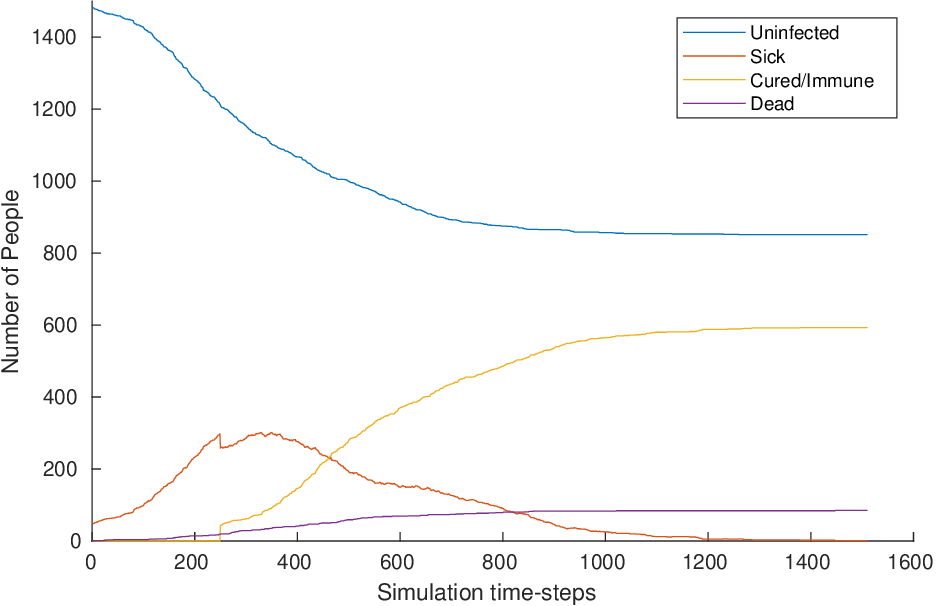}
    \caption{The simulation results for all the regions when vaccine distribution mode was "maximumUninfected".}
    \label{fig:allregionsWithVaccineMaximumUninfected}
\end{figure}
\begin{figure}[htb!]
    \centering
    \includegraphics[width=0.45\textwidth]{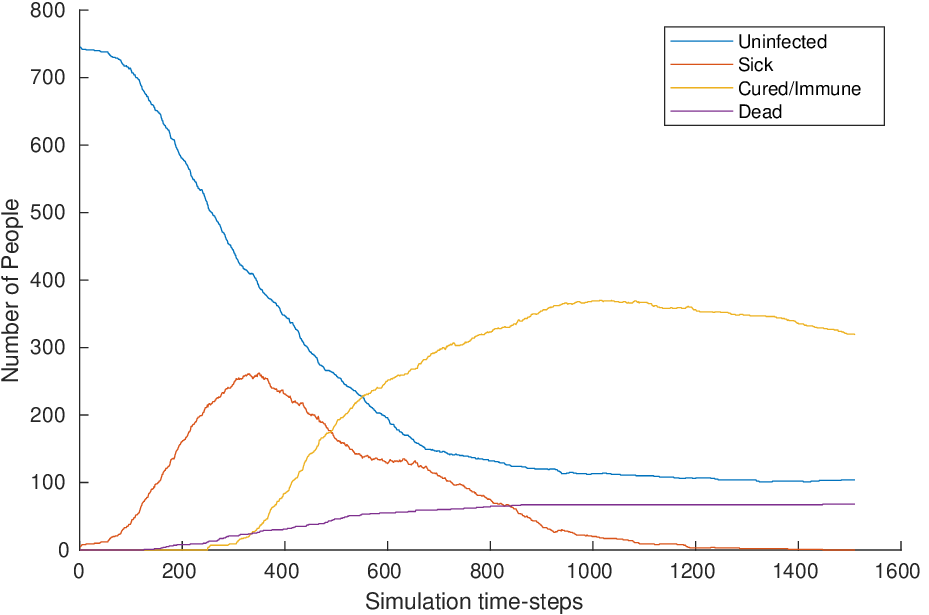}
    \caption{The simulation results for the central city when vaccine distribution mode was "maximumUninfected".}
    \label{fig:centralcityWithVaccineMaximumUninfected}
\end{figure}

At the end of the simulation, the statistics of the number of people who where dead and recovered are mentioned in Table \ref{tab:VaccineMaximumUninfectedResults}.

\begin{table}[htb!]
\centering
\begin{tabular}{|c|c|}
\hline
\textbf{Name} & \textbf{Value}  \\ \hline
Initial Population & 1530 \\ \hline
Simulation Period & 1510 \\ \hline
Total Immune & 594 \\ \hline
Total Dead & 85 \\ \hline
\end{tabular}
\caption{Tabulated final results for all the regions with vaccine distribution mode "maximumUninfected".}
\label{tab:VaccineMaximumUninfectedResults}
\end{table}

Fourthly, let us observe the results when vaccine is distributed prioritizing regions with highest sum of infected and uninfected people first. The parameter values for the same is described in Table \ref{tab:vaccineInfectedAndUninfectedParams}.

\begin{table}[htb!]
\centering
\begin{tabular}{|c|c|}
\hline
\textbf{Parameter name} & \textbf{Value}  \\ \hline
vaccineDistributionStartTime & 300 \\ \hline
vaccineDistributionFrequency & 100 \\ \hline
vaccineDistributionQuantity & 200 \\ \hline
vaccineDistributionMechanism & "infectedAndUninfected" \\ \hline
\end{tabular}
\caption{Simulation parameters used when vaccine distribution mode was "infectedAndUninfected".}
\label{tab:vaccineInfectedAndUninfectedParams}
\end{table}

\begin{figure}[htb!]
    \centering
    \includegraphics[width=0.45\textwidth]{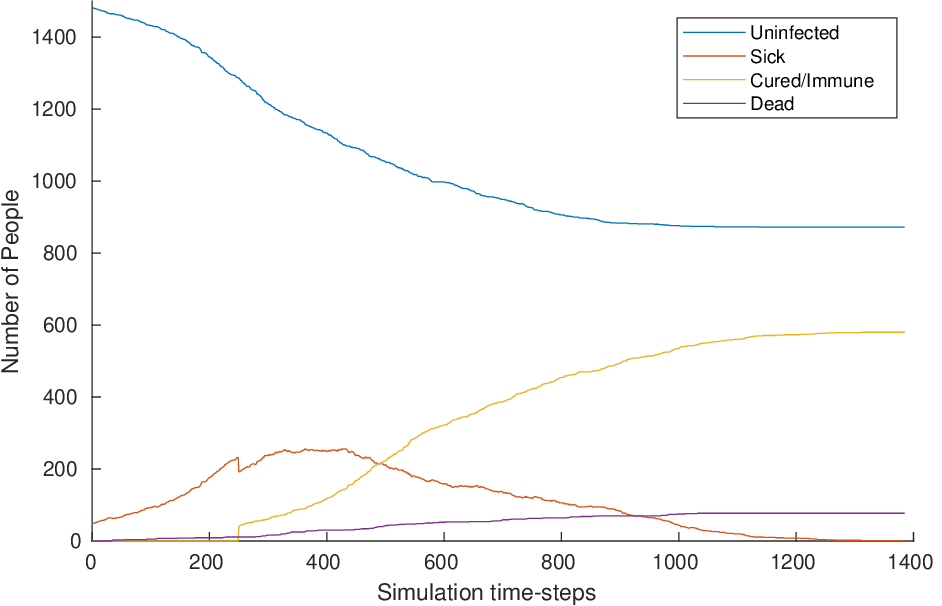}
    \caption{The simulation results for all the regions when vaccine distribution mode was "infectedAndUninfected".}
    \label{fig:allregionsWithVaccineInfectedAndUninfected}
\end{figure}
\begin{figure}[htb!]
    \centering
    \includegraphics[width=0.45\textwidth]{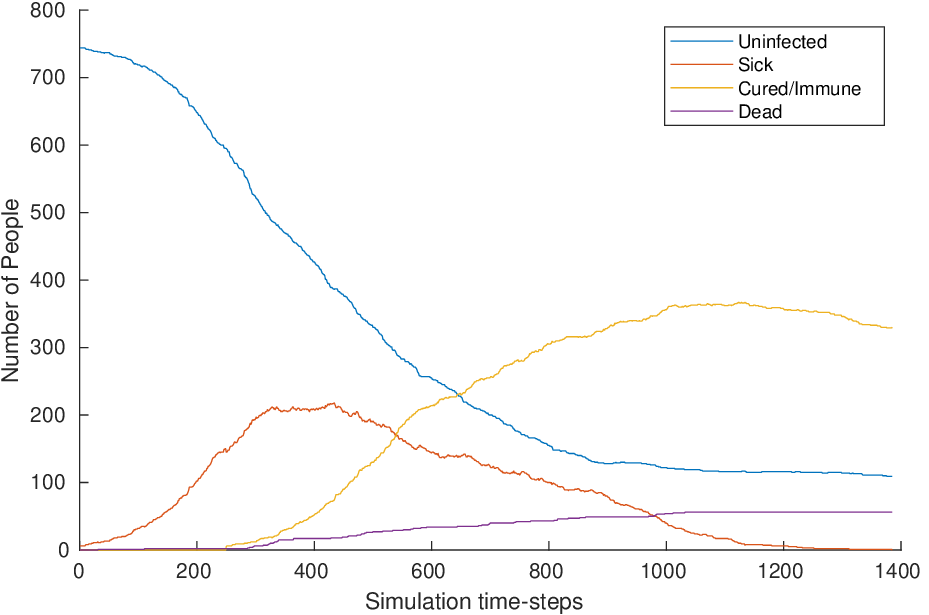}
    \caption{The simulation results for the central city when vaccine distribution mode was "infectedAndUninfected".}
    \label{fig:centralcityWithVaccineInfectedAndUninfected}
\end{figure}

At the end of the simulation, the statistics of the number of people who where dead and recovered are mentioned in Table \ref{tab:VaccineInfectedAndUninfectedResults}.

\begin{table}[htb!]
\centering
\begin{tabular}{|c|c|}
\hline
\textbf{Name} & \textbf{Value}  \\ \hline
Initial Population & 1530 \\ \hline
Simulation Period & 1385 \\ \hline
Total Immune & 581 \\ \hline
Total Dead & 77 \\ \hline
\end{tabular}
\caption{Tabulated final results for all the regions with vaccine distribution mode "infectedAndUninfected".}
\label{tab:VaccineInfectedAndUninfectedResults}
\end{table}

%%%%%%%%%%%%%%%%%%%%%%%%%%%%%%%%
\subsection{Medicine only distribution}

Firstly, let us observe the results when medicine is distributed equitably i.e., proportional to the number of uninfected people in the region at the given time. The parameter values for the same is described in Table \ref{tab:medicineEqutablyparams}.

\begin{table}[htb!]
\centering
\begin{tabular}{|c|c|}
\hline
\textbf{Parameter name} & \textbf{Value}  \\ \hline
medicineDistributionStartTime & 300 \\ \hline
medicineDistributionFrequency & 100 \\ \hline
medicineDistributionQuantity & 200 \\ \hline
medicineDistributionMechanism & "equitable" \\ \hline
\end{tabular}
\caption{Simulation parameters used when medicine distribution mode was "equitable".}
\label{tab:medicineEqutablyparams}
\end{table}

\begin{figure}[htb!]
    \centering
    \includegraphics[width=0.45\textwidth]{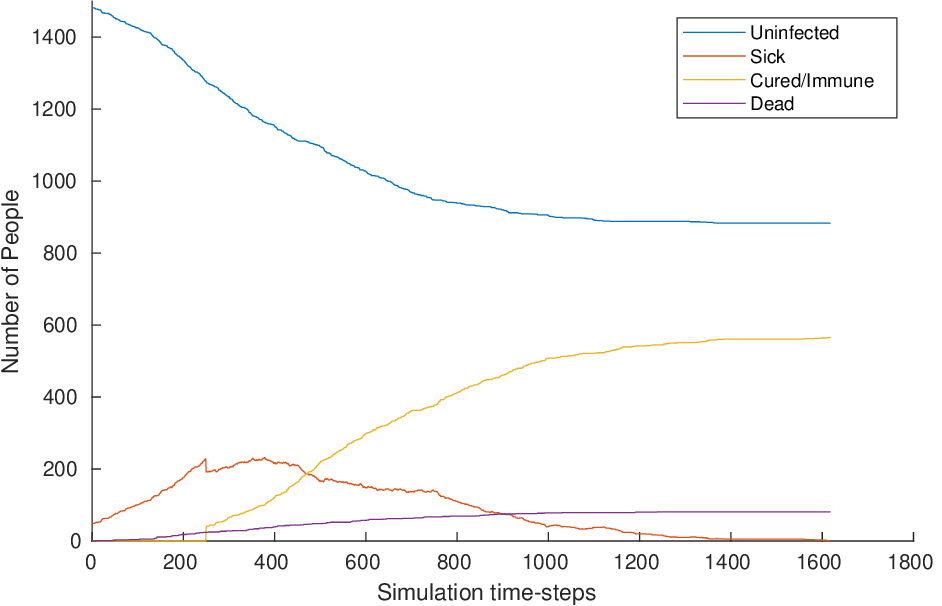}
    \caption{The aggregate simulation results for all the regions when medicine distribution mode was "equitable".}
    \label{fig:allregionsWithMedicineEqutably}
\end{figure}
\begin{figure}[htb!]
    \centering
    \includegraphics[width=0.45\textwidth]{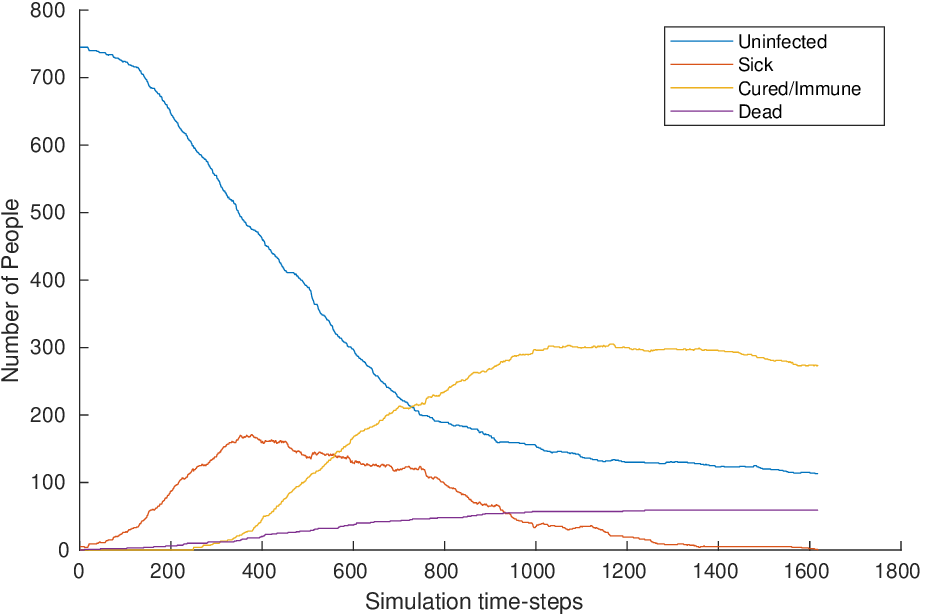}
    \caption{The simulation results for the central city when medicine distribution mode was "equitable".}
    \label{fig:centralcityWithMedicineEqutably}
\end{figure}

At the end of the simulation, the statistics of the number of people who where dead and recovered are mentioned in Table \ref{tab:medicineEqutablyResult}.

\begin{table}[htb!]
\centering
\begin{tabular}{|c|c|}
\hline
\textbf{Name} & \textbf{Value}  \\ \hline
Initial Population & 1530 \\ \hline
Simulation Period & 1618 \\ \hline
Total Immune & 566 \\ \hline
Total Dead & 81 \\ \hline
\end{tabular}
\caption{Tabulated final results for all the regions with medicine distribution mode "equitable".}
\label{tab:medicineEqutablyResult}
\end{table}

Secondly, let us observe the results when medicine is distributed prioritizing regions with maximum number of infected cases first. The parameter values for the same is described in Table \ref{tab:medicineMaximumInfectionParams}.

\begin{table}[htb!]
\centering
\begin{tabular}{|c|c|}
\hline
\textbf{Parameter name} & \textbf{Value}  \\ \hline
medicineDistributionStartTime & 300 \\ \hline
medicineDistributionFrequency & 100 \\ \hline
medicineDistributionQuantity & 200 \\ \hline
medicineDistributionMechanism & "maximumInfection" \\ \hline
\end{tabular}
\caption{Simulation parameters used when medicine distribution mode was "maximumInfection".}
\label{tab:medicineMaximumInfectionParams}
\end{table}

\begin{figure}[htb!]
    \centering
    \includegraphics[width=0.45\textwidth]{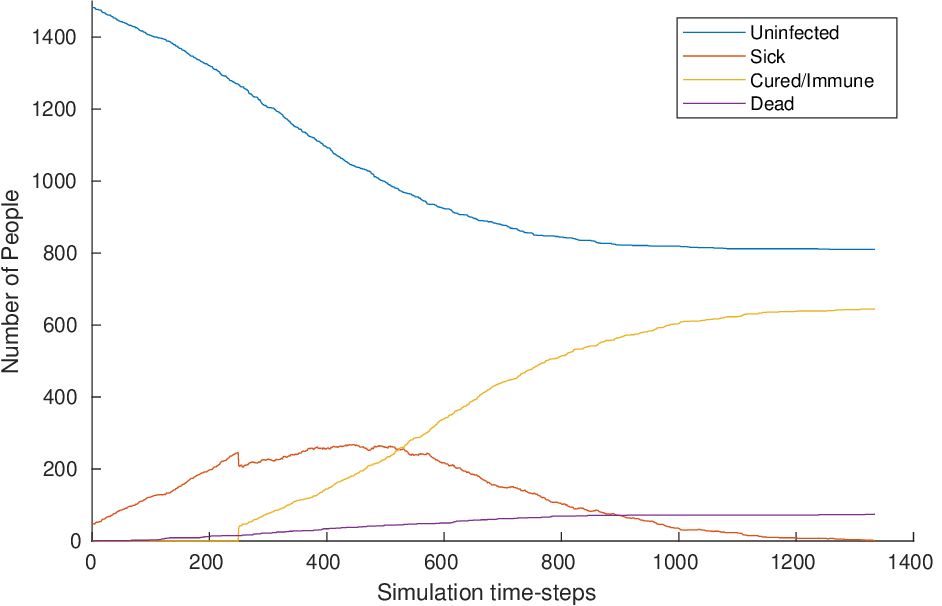}
    \caption{The simulation results for all the regions when medicine distribution mode was "maximumInfection".}
    \label{fig:allregionsWithMedicineMaximumInfection}
\end{figure}
\begin{figure}[htb!]
    \centering
    \includegraphics[width=0.45\textwidth]{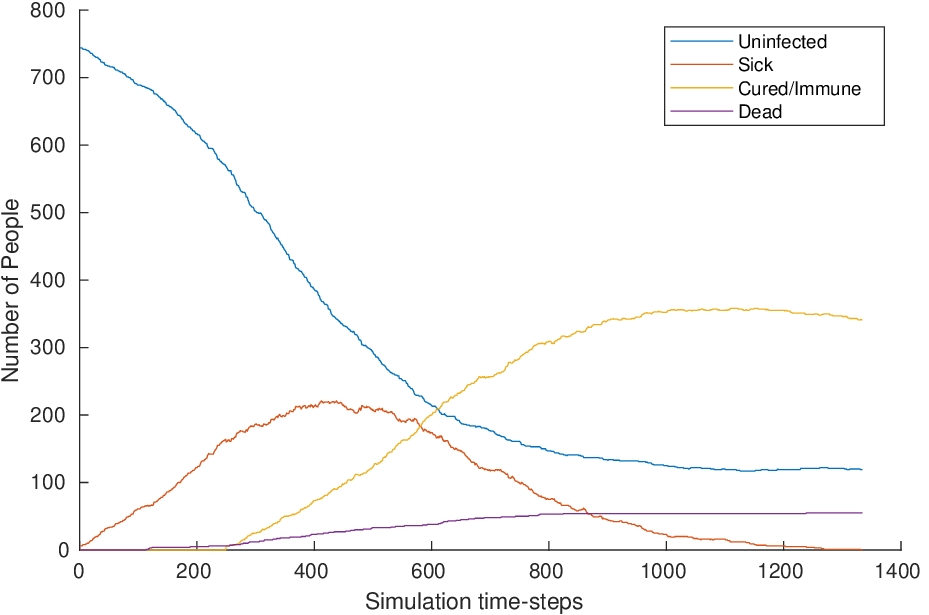}
    \caption{The simulation results for the central city when medicine distribution mode was "maximumInfection".}
    \label{fig:centralcityWithMedicineMaximumInfection}
\end{figure}

At the end of the simulation, the statistics of the number of people who where dead and recovered are mentioned in Table \ref{tab:MedicineMaximumInfectionResults}.

\begin{table}[htb!]
\centering
\begin{tabular}{|c|c|}
\hline
\textbf{Name} & \textbf{Value}  \\ \hline
Initial Population & 1530 \\ \hline
Simulation Period & 1334 \\ \hline
Total Immune & 645 \\ \hline
Total Dead & 75 \\ \hline
\end{tabular}
\caption{Tabulated final results for all the regions with medicine distribution mode "maximumInfection".}
\label{tab:MedicineMaximumInfectionResults}
\end{table}

Thirdly, let us observe the results when medicine is distributed prioritizing regions with maximum number of uninfected people first. The parameter values for the same is described in Table \ref{tab:medicineMaximumUninfectedParams}.

\begin{table}[htb!]
\centering
\begin{tabular}{|c|c|}
\hline
\textbf{Parameter name} & \textbf{Value}  \\ \hline
medicineDistributionStartTime & 300 \\ \hline
medicineDistributionFrequency & 100 \\ \hline
medicineDistributionQuantity & 200 \\ \hline
medicineDistributionMechanism & "maximumUninfected" \\ \hline
\end{tabular}
\caption{Simulation parameters used when medicine distribution mode was "maximumUninfected".}
\label{tab:medicineMaximumUninfectedParams}
\end{table}

\begin{figure}[htb!]
    \centering
    \includegraphics[width=0.45\textwidth]{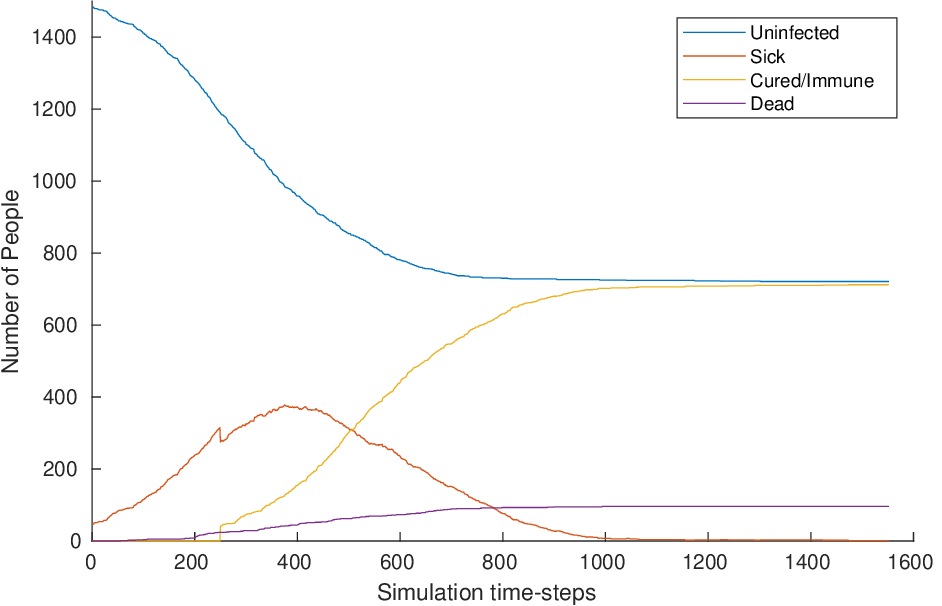}
    \caption{The simulation results for all the regions when medicine distribution mode was "maximumUninfected".}
    \label{fig:allregionsWithMedicineMaximumUninfected}
\end{figure}
\begin{figure}[htb!]
    \centering
    \includegraphics[width=0.45\textwidth]{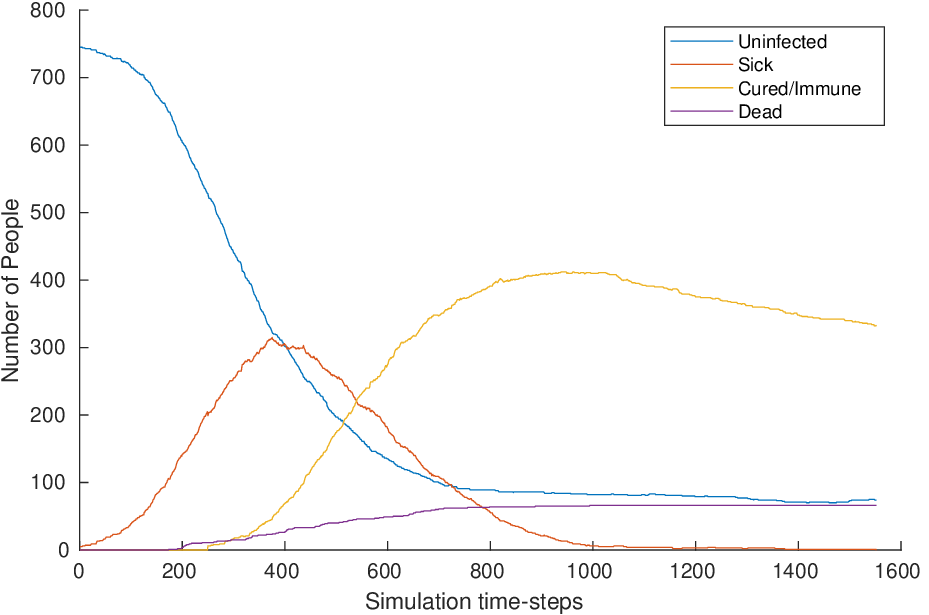}
    \caption{The simulation results for the central city when medicine distribution mode was "maximumUninfected".}
    \label{fig:centralcityWithMedicineMaximumUninfected}
\end{figure}

At the end of the simulation, the statistics of the number of people who where dead and recovered are mentioned in Table \ref{tab:MedicineMaximumUninfectedResults}.

\begin{table}[htb!]
\centering
\begin{tabular}{|c|c|}
\hline
\textbf{Name} & \textbf{Value}  \\ \hline
Initial Population & 1530 \\ \hline
Simulation Period & 1553 \\ \hline
Total Immune & 713 \\ \hline
Total Dead & 96 \\ \hline
\end{tabular}
\caption{Tabulated final results for all the regions with medicine distribution mode "maximumUninfected".}
\label{tab:MedicineMaximumUninfectedResults}
\end{table}

Fourthly, let us observe the results when medicine is distributed prioritizing regions with highest sum of infected and uninfected people first. The parameter values for the same is described in Table \ref{tab:medicineInfectedAndUninfectedParams}.

\begin{table}[htb!]
\centering
\begin{tabular}{|c|c|}
\hline
\textbf{Parameter name} & \textbf{Value}  \\ \hline
medicineDistributionStartTime & 300 \\ \hline
medicineDistributionFrequency & 100 \\ \hline
medicineDistributionQuantity & 200 \\ \hline
medicineDistributionMechanism & "infectedAndUninfected" \\ \hline
\end{tabular}
\caption{Simulation parameters used when medicine distribution mode was "infectedAndUninfected".}
\label{tab:medicineInfectedAndUninfectedParams}
\end{table}

\begin{figure}[htb!]
    \centering
    \includegraphics[width=0.45\textwidth]{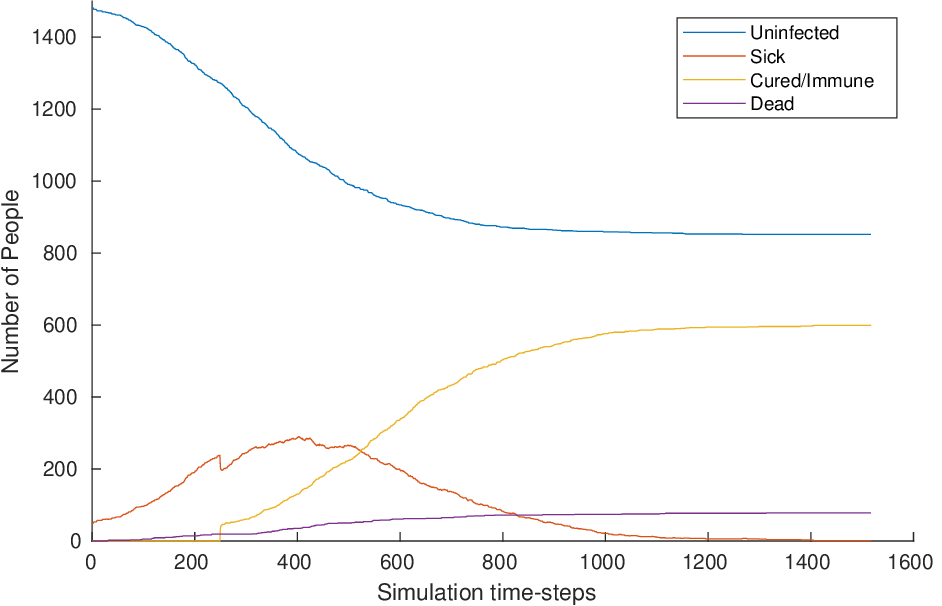}
    \caption{The simulation results for all the regions when medicine distribution mode was "infectedAndUninfected".}
    \label{fig:allregionsWithMedicineInfectedAndUninfected}
\end{figure}
\begin{figure}[htb!]
    \centering
    \includegraphics[width=0.45\textwidth]{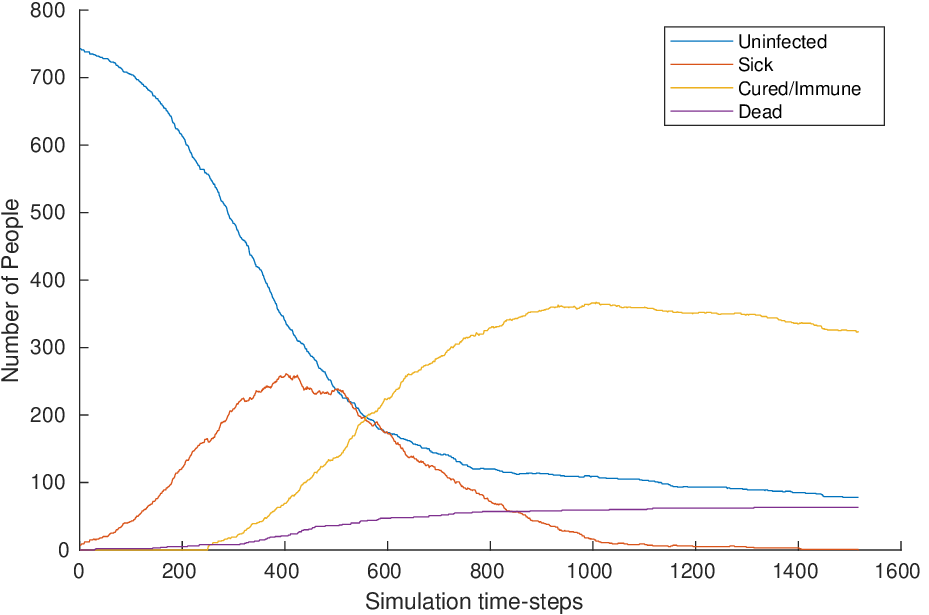}
    \caption{The simulation results for the central city when medicine distribution mode was "infectedAndUninfected".}
    \label{fig:centralcityWithMedicineInfectedAndUninfected}
\end{figure}

At the end of the simulation, the statistics of the number of people who where dead and recovered are mentioned in Table \ref{tab:MedicineInfectedAndUninfectedResults}.

\begin{table}[htb!]
\centering
\begin{tabular}{|c|c|}
\hline
\textbf{Name} & \textbf{Value}  \\ \hline
Initial Population & 1530 \\ \hline
Simulation Period & 1517 \\ \hline
Total Immune & 600 \\ \hline
Total Dead & 78 \\ \hline
\end{tabular}
\caption{Tabulated final results for all the regions with medicine distribution mode "infectedAndUninfected".}
\label{tab:MedicineInfectedAndUninfectedResults}
\end{table}

%%%%%%%%%%%%%%%%%%%%%%%%%%%%%%%%
\subsection{Simultaneous Vaccine and Medicine distribution}

With medicine and vaccine distribution starting at 300 time-step with a quantity of 200 units and getting replenished after every 100 time-steps -Vaccine prioritizing regions with maximum number of the sum of uninfected and infected people first and medicine prioritizing regions with maximum number of infected people first - Lockdown enforced when more than 10\% infected and revoked when less than 2\% infected. The parameter values for the same is described in Table \ref{tab:LockdownMedicineVaccineParams}.

\begin{table}[htb!]
\centering
\begin{tabular}{|c|c|}
\hline
\textbf{Parameter name} & \textbf{Value}  \\ \hline
lockdownStartThreshold & 0.1 \\ \hline
lockdownEndThreshold & 0.02 \\ \hline
lockdownMobilityMultiplier & 0.1 \\ \hline
vaccineDistributionStartTime & 300 \\ \hline
vaccineDistributionFrequency & 100 \\ \hline
vaccineDistributionQuantity & 200 \\ \hline
vaccineDistributionMechanism & "infectedAndUninfected" \\ \hline
medicineDistributionStartTime & 300 \\ \hline
medicineDistributionFrequency & 100 \\ \hline
medicineDistributionQuantity & 200 \\ \hline
medicineDistributionMechanism & "maximumInfection" \\ \hline
\end{tabular}
\caption{Simulation parameters used when medicine distribution mode was "infectedAndUninfected".}
\label{tab:LockdownMedicineVaccineParams}
\end{table}

\begin{figure}[htb!]
    \centering
    \includegraphics[width=0.45\textwidth]{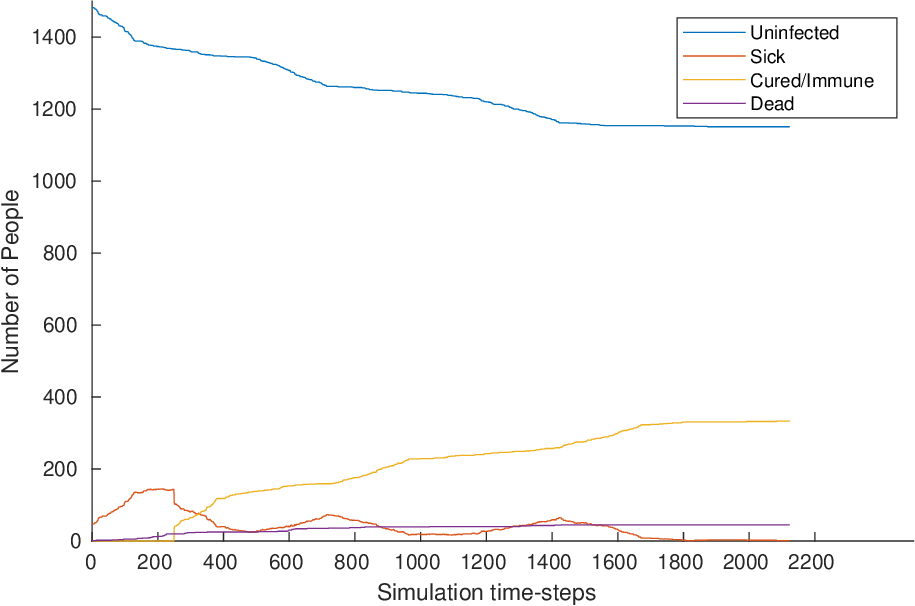}
    \caption{The simulation results for all the regions with lockdown imposition, and, vaccine and medicine distribution}
    \label{fig:allregionsWithLockdownMedicineVaccine}
\end{figure}
\begin{figure}[H]
    \centering
    \includegraphics[width=0.45\textwidth]{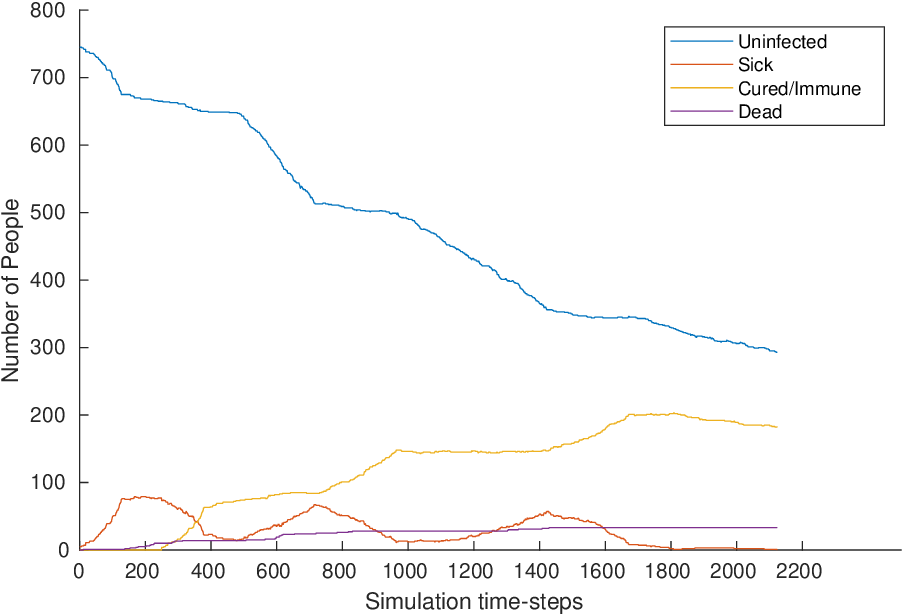}
    \caption{The simulation results for  the central city with lockdown imposition, and, vaccine and medicine distribution}
    \label{fig:centralcityWithLockdownMedicineVaccine}
\end{figure}

At the end of the simulation, the statistics of the number of people who where dead and recovered are mentioned in Table \ref{tab:LockdownMedicineVaccineResults}.

\begin{table}[htb!]
\centering
\begin{tabular}{|c|c|}
\hline
\textbf{Name} & \textbf{Value}  \\ \hline
Initial Population & 1530 \\ \hline
Simulation Period & 2123 \\ \hline
Total Immune & 334 \\ \hline
Total Dead & 45 \\ \hline
\end{tabular}
\caption{Tabulated final results for all the regions with medicine distribution mode "infectedAndUninfected".}
\label{tab:LockdownMedicineVaccineResults}
\end{table}

%%%%%%%%%%%%%%%%%%%%%%%%%%%%%%%%%%%%%%%%%%%%%%%%%%%%%%%%%%%%%%%%%%%%%%%%%%%%%%%%
\section{Conclusions}

The designed framework is able to simulate wide variety of scenarios. Indeed, there are infinitely many possible combinations of vaccine and medicine distribution that can be, and some of these were explored in this article. Based on different social and economic conditions, the decision makers may tweak different parameters to virtually test their decision making skills in the future using this framework. However, do note that the framework does have some assumptions, and it may be improved in the future to have less of them.

\bibliography{aaai23}

\newpage
%%%%%%%%%%%%%%%%%%%%%%%%%%%%%%%%%%%%%%%%%%%%%%%%%%%%%%%%%%%%%%%%%%%%%%%%%%%%%%%%
\section{Appendix}

\subsection{Code and data}

The implementation code, generated data, results and analysis are available in \url{https://github.com/zenineasa/frameworkForVaccineAndCureDistribution}.

%%%%%%%%%%%%%%%%%%%%%%%%%%%%%%%%
\subsection{Hardware and software setup}

The specifications for the device and the software libraries used to perform the simulations are described in Table \ref{tab:linuxSpec} and Table \ref{tab:softwareVersions}. However, note that the framework developed to run on any modern computer using a modern browser, irrespective of the operating system. Although we used MATLAB for analysing and interpreting the results, one could alternatively use Python or Excel.

\begin{table}[H]
    \centering
    \begin{tabular}{|p{0.12\textwidth}|p{0.30\textwidth}|}
        \hline
        \textbf{Component} & \textbf{Specification} \\
        \hline
        Hardware & Lenovo ideapad 330-15ARR \\
        Processor & AMD Ryzen 5 2500U \\
        No. of cores & 8 \\
        Graphics & AMD Radeon vega 8 graphics \\
        Memory & 8 GB ($2 \times$ 4 GB DDR4 @ 2400 MHz)\\
        OS & Ubuntu 20.04.02 LTS \\
        \hline
    \end{tabular}
    \caption{Specification of the machine used for performance evaluation}
    \label{tab:linuxSpec}
\end{table}

\begin{table}[H]
    \centering
    \begin{tabular}{|p{0.16\textwidth}|p{0.26\textwidth}|}
        \hline
        \textbf{Software / Library} & \textbf{Version} \\
        \hline
        Google Chrome & 86.0.4240.183 \\
        MATLAB & R2020b \\
        \hline
    \end{tabular}
    \caption{Versions of different software and libraries used}
    \label{tab:softwareVersions}
\end{table}

%%%%%%%%%%%%%%%%%%%%%%%%%%%%%%%%
\subsection{Choosing to build a new framework}

The reason for building a framework that can run on a modern browser was motivated by the fact that these are free to use, have immense capabilities, and are nearly universally accessible. Decision makers need not require programmers to try different strategies and observe the potential outcome of their decisions. Ease of use and access were our primary priority while developing this framework.

\end{document}